\documentstyle[12pt,aps,graphicx,psfrag,float]{revtex}

\begin{document}

\draft
\title{Thermodynamics of toroidal black holes}
\author{Claudia S. Pe\c ca\footnote{Electronic address:
    claudia@feynman.ist.utl.pt}} 
\address{Departamento de F\' \i sica, Instituto Superior T\'ecnico,
  Av. Rovisco Pais 1, 1096 Lisboa Codex, Portugal} 
\author{Jos\'e P. S. Lemos\footnote{Electronic address: lemos@on.br}}
\address{Departamento de Astrof\' \i sica. Observat\'orio
  Nacional-CNPq, Rua General Jos\'e Cristino 77, 20921 Rio de Janeiro,
  Brazil \\ and 
Departamento de F\' \i sica, Instituto Superior T\'ecnico,
  Av. Rovisco Pais 1, 1096 Lisboa Codex, Portugal} 
\date{\today}
\maketitle
\begin{abstract}
The thermodynamical properties of toroidal black holes in the grand
canonical ensemble are investigated using York's formalism. The black
hole is enclosed in a cavity with finite radius where the temperature
and electrostatic potential are fixed. The boundary conditions allow
one to compute the relevant thermodynamical quantities, e.g. thermal
energy, entropy and specific heat. This black hole is
thermodynamically stable and dominates the grand partition
function. This means that there is no phase transition, 
as the one encountered for spherical black holes. 
\end{abstract}

\newpage

\section{Introduction}

In recent years there has been an increasing interest in the study of
black holes with non-trivial topologies. Black holes whose event
horizon have toroidal topology have been found, see
\cite{Lemos95,Zanchin}. It has  been shown that they can
be formed from  gravitational collapse \cite{Lemos98}.  There is also a
generalization of these black holes to other topologies
\cite{Amin,Brill}. 
These black holes are solutions of the Einstein equations with negative
cosmological constant. It is the  presence of a negative cosmological
constant that allows the violation of the theorems of general
relativity forbidding non-spherical black hole topologies. 

It is interesting to study the thermodynamical properties of these
black holes and compare them to their spherically symmetric
counterparts. In this paper we consider  a static and charged black
hole with toroidal event horizon found in \cite{Zanchin} and analyze
its thermodynamical behaviour, see also \cite{Brill,Vanzo}.  We use York's
formalism to study its thermodynamics in the grand canonical following
the same procedure as in \cite{rnads} for its spherical counterpart, the
Reissner-Nordstr\"om-anti-de~Sitter black hole.

In section \ref{therm} we compute the reduced action of the toroidal
black hole in York's formalism. Using this action we evaluate the main
thermodynamical quantities: energy, mean value of the charge and
entropy of the toroidal black hole. In section \ref{sol} we evaluate
the black hole solutions, i.e. the event horizon radius and charge of
the black hole formed for the temperature and  the electrostatic
potential fixed by the boundary conditions. 
We then study the local and global stability of the black hole
solutions in section \ref{stab}. Finally, in section \ref{infty} we
consider the limit where the boundary is taken to
infinite. The results obtained are compared to the ones found in
\cite{rnads} for the Reissner-Nordstr\"om-anti-de~Sitter black
hole. Conclusions are drawn in section \ref{concl}.

\section{Thermodynamics in the grand canonical ensemble}
\label{therm}

In this section we will compute the reduced action of the black hole. 
We will follow \cite{Braden}.

We consider a general static metric with toroidal symmetry of the form
\begin{equation}
  \label{metric}
  ds^2=b^2 d\tau^2+a^2 dy^2+r^2 (d\theta^2 + d\varphi^2)\ ,
\end{equation}
where $a$, $b$, and $r$ are only function of the radial coordinate
$y$. The Euclidean time $\tau$ and the angular coordinates $\theta$
and $\varphi$  have period $2 \pi$. 
 For convenience we choose $y\in[0,1]$ so that the event 
horizon is given by $y=0$ and has radius $r_+=r(0)$ and area $A_+=4 \pi^2
r_+^2$. The boundary is given by $y=1$ and at this boundary  the
thermodynamical variables  defining the ensemble are fixed. The
boundary is a 2-torus with area $A_B=4 \pi^2 r_B^2$, where $r_B=r(1)$.

In order to obtain the reduced action  from (\ref{metric})
we use the usual regularity conditions  and we impose the proper
constraints \cite{Braden,York89}, i.e. Hamiltonian constraint 
\begin{equation}
  \label{hamconstr}
  {G^\tau}_\tau + \Lambda {g^\tau}_\tau = 8 \pi {T^\tau}_\tau \ ,
\end{equation}
which corresponds to the first of Einstein equations, 
and the Gaussian constraint 
\begin{equation}
  \label{gaussconstr}
  {F^{\mu\nu}}_{;\nu} = 0 \ ,
\end{equation}
which corresponds to the Maxwell equations. 

The reduced action is given by
\begin{equation}
  \label{I^*}
  I^* = - \beta \, r_B \sqrt{ \pi^2 \alpha^2 r_B^2 - \pi^2 \alpha^2
    \frac{r_+^3}{r_B} - \frac{e^2}{r_+\, r_B} + \frac{e^2}{r_B^2}} -
  \pi^2 r_+^2 - e \, \beta \, \phi - I_{\rm subtr} \ .
\end{equation}
Where $\beta$ is the inverse temperature at the boundary, $\phi$ is the
difference in electrostatic potential between the boundary and the
horizon, $e$ the charge of the black hole, $\alpha^2 =
-\frac{\Lambda}{3}$, $\Lambda$ is the cosmological constant and
$I_{\rm subtr}$ is an arbitrary term that  can be used to define the
zero of the energy. 

Using the same procedure as in \cite{Braden,rnads}, we can compute
$I_{\rm subtr}$. We choose for convenience the thermal energy of
anti-de~Sitter spacetime $E_{\rm ADS}=E(r_+=0,e=0)=0$ to
define the zero of the energy. Therefore obtaining 
\begin{equation}
  \label{Isubtr}
  I_{\rm subtr}= \beta r_B \sqrt{\pi^2 \alpha^2 r_B^2} \ .
\end{equation}
Now substituting (\ref{Isubtr}) in (\ref{I^*}) we obtain the reduced
action in the form 
\begin{equation}
  \label{I*}
  I^* =  \beta \, r_B \left(\sqrt{\pi^2 \alpha^2 r_B^2} - \sqrt{ \pi^2
      \alpha^2 r_B^2 - \pi^2 \alpha^2 \frac{r_+^3}{r_B} -
      \frac{e^2}{r_+\, r_B} + \frac{e^2}{r_B^2}} \right) - \pi^2 r_+^2
  - e \, \beta \, \phi  \ .
\end{equation}

We can use the reduced action given in (\ref{I*}) to compute all the
thermodynamical quantities of interest (like the energy, the entropy
and the mean value of charge of the black hole). This is done using
the relation between the reduced action and the grand canonical
potential of thermodynamics \cite{Braden}
\begin{equation}
  \label{IBF}
  I = \beta F \ .
\end{equation}
>From the grand canonical potential $F$, we can compute the thermodynamical
quantities using the common laws of thermodynamics, see for
example \cite{callen}. 

The thermal energy is given by
\begin{eqnarray}
\label{E}
E &=& F+ \beta \left( \frac{\partial F}{\partial
      \beta}\right)_{\phi,r_B} - \left( \frac{\partial F}{\partial
      \phi} \right)_{\beta,r_B} \phi  =
\left( \frac{\partial I}{\partial \beta}\right)_{\phi,r_B}-
\frac{\phi}{\beta} \left( \frac{\partial I}{\partial \phi}
\right)_{\beta,r_B} = \nonumber \\ &=&  r_B \left(\sqrt{\pi^2 \alpha^2
    r_B^2} - \sqrt{ \pi^2 \alpha^2 r_B^2 - \pi^2 \alpha^2
    \frac{r_+^3}{r_B} - \frac{e^2}{r_+\, r_B} + \frac{e^2}{r_B^2}}
\right) \ .
\end{eqnarray}

The mean value of the charge is given by
\begin{equation}
  \label{Q}
  Q = -\left( \frac{\partial F}{\partial \phi} \right)_{\beta,r_B} = 
- \frac{1}{\beta} \left( \frac{\partial I}{\partial \phi}
\right)_{\beta,r_B} = e \ .
\end{equation}

The entropy is 
\begin{equation}
  \label{S}
  S = \beta^2 \left(\frac{\partial F}{\partial \beta}\right)_{\phi,r_B}=
\beta \  \left(\frac{\partial I}{\partial \beta}\right)_{\phi,r_B}
  - I = \pi^2 r_+^2\ .
\end{equation}
Since $\pi^2 \, r_+^2= A_+/4$, where $A_+$ is the area of the event
horizon, we have $S=\frac{A_+}{4}$. This is the usual
Hawking-Bekenstein entropy \cite{Hawk75}, which means  this law
is still valid for black holes with toroidal symmetry.  

\section{The black hole solutions}
\label{sol}

The black hole solutions are determined by computing the extrema of
the reduced action. As the variables $\beta$, $\phi$, $r_B$ and
$\alpha$ are fixed by the boundary conditions, the reduced action
(\ref{I*}) is a function of only two parameters: $r_+$ the event
horizon radius and $e$ the electric charge. Inverting the equation
$\nabla I^*(r_+,e)=0$, we obtain the black hole solutions as function
of the boundary conditions, i.e. $r_+=r_+(\beta,\phi,r_B,\alpha)$ and
$e=e(\beta,\phi,r_B,\alpha)$. Equation $\nabla I^*=0$ yields
\begin{equation}
  \label{dir+}
  \frac{\partial I^*}{\partial r_+} = -\frac{1}{2} \, \beta \,
  \left(-3 \, \pi^2 \, \alpha^2 \, r_+^2 + \frac{e^2}{r_+^2} \right)
  \left(\pi^2 \, \alpha^2 \, r_B^2- \pi^2 \, \alpha^2 \,
    \frac{r_+^3}{r_B}- \frac{e^2}{r_+ \, r_B}+\frac{e^2}{r_B^2}
  \right)^{-\frac{1}{2}} -2 \, \pi^2 \, r_+ = 0 \ , 
\end{equation}
and
\begin{equation}
  \label{die}
  \frac{\partial I^*}{\partial r_+} = - \beta \ \left(- \frac{e}{r_+}
    + \frac{e}{r_B} \right) \left(\pi^2 \, \alpha^2 \, r_B^2- 
      \pi^2 \, \alpha^2 \, \frac{r_+^3}{r_B}- \frac{e^2}{r_+ \,
        r_B}+\frac{e^2}{r_B^2}\right)^{-\frac{1}{2}} - \beta \, \phi = 0 \ .
\end{equation}

We can invert equation (\ref{dir+}) to obtain the inverse temperature
of the black hole
\begin{equation}
  \label{beta}
  \beta = \frac{4 \, \pi^2 \, r_+^3}{3 \, \pi^2 \, \alpha^2 \,
    r_+^4-e^2} \sqrt{\pi^2 \, \alpha^2 \, r_B^2-
      \pi^2 \, \alpha^2 \, \frac{r_+^3}{r_B}- \frac{e^2}{r_+ \,
        r_B}+\frac{e^2}{r_B^2}} \ .
\end{equation}
This  is the Hawking temperature times the redshift factor due to
the Tolman effect \cite{Tolman}. 

Inverting  equation (\ref{die}), one obtains the electrostatic
potential as
\begin{equation}
  \label{phi}
  \phi = \left( \frac{e}{r_+} - \frac{e}{r_B} \right) \left(\pi^2 \,
    \alpha^2 \, r_B^2- \pi^2 \, \alpha^2 \, \frac{r_+^3}{r_B}-
    \frac{e^2}{r_+ \, r_B}+\frac{e^2}{r_B^2}\right)^{-\frac{1}{2}} \ .
\end{equation}
This is the difference in electrostatic potential between the horizon
and the boundary ``redshifted'' to the boundary. 

In order to invert equations (\ref{beta}) and (\ref{phi}), we are
going to define the new variables 
\begin{equation}
  \label{new}
  \overline \alpha = \pi \, \alpha \, r_B \ \ , \ \ \ x =
  \frac{r_+}{r_B} \ \ , \ \ \ q = \frac{e}{r_B} \ \ , \ \ \ \overline
  \beta = \frac{\beta}{4 \, \pi^2 \, r_B} \ .
\end{equation}

Using these new variables and inverting equation (\ref{phi}) we obtain
\begin{equation}
  \label{q}
  q^2 = \frac{\overline \alpha^2 \, \phi^2 \, (1+x+x^2) \, x^2
    }{1-x+\phi^2 \, x} \ .
\end{equation}
Inverting equation (\ref{beta}) and using equation (\ref{q}), we
obtain the equation 
\begin{eqnarray}
  \label{x^5}
  \sigma^2 &\, \phi^4 &+ \, 4\, \sigma^2 \, \phi^4 \, x+ (10 \, \sigma^2 \,
  \phi^4-6 \, \sigma^2 \, \phi^2 -1)\, x^2 + \phi^2 \, (12 \,\sigma^2 \,
  \phi^2 -12 \, \sigma^2 -1)\, x^3 \nonumber \\ {}&+& (9 \, \sigma^2
  \, \phi^4 -18 \, \sigma^2 \, \phi^2 - 9 \, \sigma^2 -\phi^2) \, x^4
  + (1-\phi^2)\, x^5 = 0 \ .
\end{eqnarray}
Where we have used a new variable
\begin{equation}
  \label{sigma}
  \sigma = \overline \alpha \, \overline \beta =
    \frac{\alpha \, \beta}{4\,\pi} \ . 
\end{equation}

Notice that form equation (\ref{x^5}) the event horizon radius does
not depend on $\beta$ and $\alpha$, but on their product
$\sigma$. Something that does not happen for the
Reissner-Nordstr\"om-anti-de~Sitter black hole \cite{rnads}, the
spherical counterpart of this black hole.

\begin{figure}
  \begin{center}
\psfrag{x}{$x$}
\psfrag{s}{$\sigma$}
\psfrag{f}{${\phi}$}
    \includegraphics[height=5cm,width=10cm]{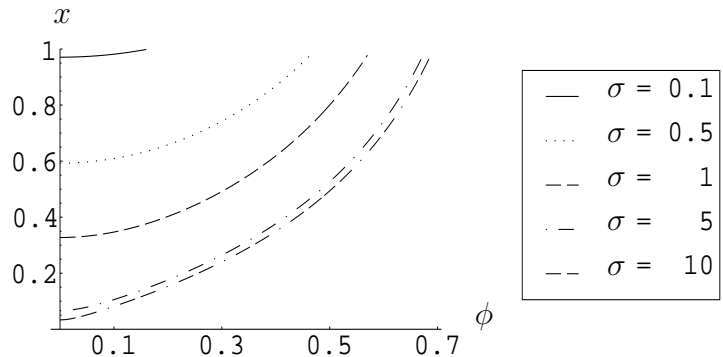}
    \caption{Solutions of equation (\ref{x^5}) which  obey condition
      (\ref{f<}), for fixed values of $\sigma=$ 0.1, 0.5, 1, 5, 10.} 
    \label{fig1}
  \end{center}
\end{figure}

\begin{figure}
  \begin{center}
\psfrag{x}{$x$}
\psfrag{s}{$s$}
\psfrag{f}{$\phi$}
    \includegraphics[height=7cm,width=7cm]{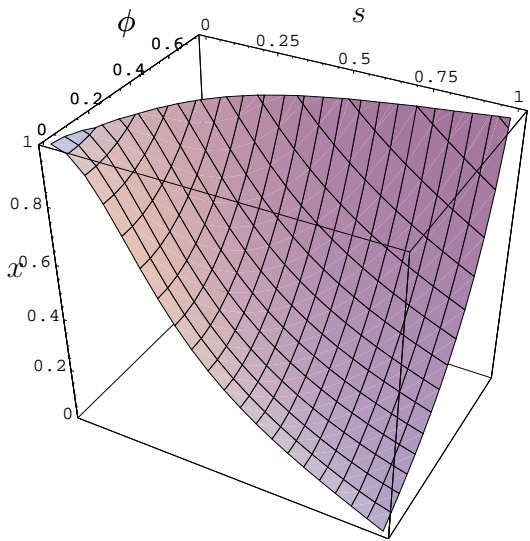}
    \caption{Solutions of equation (\ref{x^5}) which  obey condition
      (\ref{f<}). The variable $s$ is defined in (\ref{ss}).} 
    \label{fig3}
  \end{center}
\end{figure}

Solving equation (\ref{x^5}), we obtain $r_+$ as a function of the
boundary conditions and replacing this solution into equation
(\ref{q}), we obtain the respective charge $e$. However not every
solution of equation (\ref{x^5}) is a physical solution corresponding
to a black hole. Effectively, the black hole is inside the cavity so
$r_+< r_B$, therefore the solutions must obey $x<1$. Moreover the
charged black hole has two horizons but we are only interested in the
event horizon, which verifies condition
\begin{equation}
  \label{event}
  3 \, \pi^2 \, \alpha^2 \, r_+^4 - e^2 > 0 \ .
\end{equation}
This condition implies the inverse temperature (\ref{beta}) is real
and positive and that the electrostatic potential (\ref{phi}) is also
real and positive and furthermore verifies
\begin{equation}
  \label{f<}
  \phi^2 < \frac{3 \, x^2}{1+2\,x+3\,x^2} \ .
\end{equation}
Therefore only the solutions of equation (\ref{x^5}) that obey
condition (\ref{f<}) are physical solutions. 
In figures \ref{fig1} and \ref{fig3}  the solutions of (\ref{x^5}) that
verify this condition are presented. 


In figure \ref{fig1}, the curves have fixed values of the variable
$\sigma$ and the values of $x$ are presented as function of the
electrostatic potential. 
Notice that, due to condition (\ref{f<}), the electrostatic potential
is always $\phi<\sqrt{0.5}\simeq .7$. 

In order to present in graphics all possible values of $\sigma$, we
define the new variable
\begin{equation}
  \label{ss}
  s = \frac{2}{\pi} \, \arctan{ \sigma } \ .
\end{equation}
It is this new variable that is used in figure \ref{fig3}, where the
black hole solutions are again presented as functions of $\sigma$ and
$\phi$. 


\section{Stability}
\label{stab}

To study the stability of the solutions found in the previous section
we compute the local minima of the reduced action (\ref{I*}). We will
follow the same procedure as \cite{Braden}. The conditions of local
stability are, see \cite{rnads},
\begin{equation}
\label{est}
   \beta \left(\frac{\partial \phi}{\partial e}\right)_{S,r_B} \geq 0  \ ,
\end{equation}
\begin{equation}
 C_{\phi,r_B} = \beta \left(\frac{\partial S}{\partial
   \beta}\right)_{\phi,r_B} \geq 0 \ ,
\end{equation}
where $C_{\phi,r_B}$ is the heat capacity at constant $\phi$ and
$r_B$. Computing these functions we obtain
\begin{equation}
  \label{piv1}
  \beta \left(\frac{\partial \phi}{\partial e}\right)_{S,r_B} = 
  \frac{4 \, \pi^4 \, \alpha^2 \, (r_B^3-r_+^3) \, r_+^3}{(3 \, \pi^2 \,
    \alpha^2 \, r_+^4 -e^2)(\pi^2 \, \alpha^2 \, r_+ \, r_B \,
    (r_B^2+r_+\,r_B+r_+^2)-e^2)} \geq 0
\end{equation}
and
\begin{equation}
  \label{piv2}
  C_{\phi,r_B} = \frac{4 \, \pi^4 \, \alpha^2 r_+^3 \, (r_B^3-r_+^3)
    \, (3\,\pi^2 \, \alpha^2 r_+^4-e^2)}{e^4 + 2 \pi^2 \alpha^2 e^2
    r_+ r_B \, (r_B^2 -2\, r_+\,r_B -2 \, r_+^2)+ 3 \pi^4 \, \alpha^4
    r_+^5 (2\,r_B^3+r_+^3)} \geq 0
\end{equation}
These conditions are satisfied for every value of $r_+$ and $e$ that
verify conditions $r_+<r_B$ and (\ref{event}). Therefore all physical
black hole solutions are locally stable. This means that the toroidal
black holes are more stable than the spherical ones, since the
Reissner-Nordstr\"om-anti-de~Sitter black hole is unstable for a wide
regions of values of $\beta$, $\phi$ and $\alpha$ \cite{rnads}.

However these solutions are not necessarily global minima of the
reduced action. In this case they do not dominate the grand partition
function and the zero-loop approximation being used here does not
hold \cite{WhitYork}. 

The reduced action given in (\ref{I*}) goes to infinity in the
non-compact directions where $r_+$ or $e$ go to infinity. Therefore the
global minimum of the reduced action is either at the local minimum or
at $r_+=e=0$. At this latter point the reduced action is
null. Therefore the condition for global stability of
the solutions computed in the previous section is that the classical
action, i.e. the reduced action evaluated at the local minimum, is
negative. This is indeed the case for all physical solutions of
equation (\ref{x^5}). We conclude that the solutions
presented in figures \ref{fig1} and \ref{fig3} are globally stable and
dominate the grand partition function. Again, we can say that the
toroidal black hole is more stable than its spherical counterpart, the
Reissner-Nordstr\"om-anti-de~Sitter black hole, which is not dominant
in a certain region of values of $\beta$, $\phi$ and $\alpha$
\cite{rnads}.  

\section{Taking the boundary to infinity}
\label{infty}

As for the spherical counterpart of this black hole \cite{rnads}, there
are two ways of taking the limit $r_B \to \infty$, ($i$) fixing
the black hole solutions, i.e. fixing the values of $r_+$ and $e$ and
($ii$) fixing the boundary conditions, i.e. fixing the values of
$\beta$ and $\phi$.  

Fixing the black hole solutions and taking the limit $r_B\to\infty$,
the temperature $T=\beta^{-1}$ and 
the electrostatic potential go to zero as $r_B^{-1}$, see equations
(\ref{beta}) and (\ref{phi}). In this case the classical action, is
given by, see equation (\ref{I*}),
\begin{equation}
  \label{Iinfty}
  I= \frac{\pi^2 \, r_+^2\, (e^2+\pi^2 \, \alpha^2 \, r_+^4)}{e^2-3\,
    \pi^2\, \alpha^2\, r_+^4}\ .
\end{equation}
This is always negative as long as the $r_+$ obeys the necessary
condition to represent the event horizon, i.e. condition
(\ref{event}). Therefore the locally stable solutions are also
globally stable and dominate the grand partition function. This means
that for this black hole there is no phase transition as the one found
for spherical black holes \cite{HP83,rnads}.
The thermal energy goes to zero as $\frac{m}{\pi \, \alpha \,
  r_B}$, where $m$ is the mass of the black hole given by
\begin{equation}
  \label{mass}
   m= \frac{\pi}{2}\left( \frac{e^2}{\pi^2 r_+} + \alpha^2 r_+^3
   \right) \ .
\end{equation}

The heat capacity is given by, see equation (\ref{piv2}),
\begin{equation}
  \label{cinfty}
  C_{\phi}= 2 \, \pi^2 \, r_+^2 \ \left( 1- \frac{2\, e^2}{e^2+3 \,
      \pi^2 \, \alpha^2 \, r_+^4}\right) \ .
\end{equation}
The heat capacity is positive as long as $r_+$ obeys condition
(\ref{event}), which means these solutions are all stable.

Fixing the boundary conditions and taking the limit $r_B \to \infty$,
we obtain solutions that diverge. This can be seen using equation
(\ref{x^5}) and taking this limit, the event horizon radius $r_+$ goes
to infinity as $x\,r_B$. All other thermodynamical quantities: mean
value of the charge, entropy, action, energy and heat capacity (see
equations (\ref{q}), (\ref{S}), (\ref{I*}), (\ref{E}) and
(\ref{piv2}), respectively), diverge as $r_B^2$. Therefore this way of
taking the limit seems to be of less physical interest than the previous. 

\section{Conclusions}
\label{concl}

We have studied the thermodynamics of the charged, static and toroidal
black hole (studied in \cite{Zanchin}) closed in a box with finite
radius. We conclude that the Hawking-Bekenstein law for entropy is
still valid for black holes with this symmetry. Furthermore we find
that in the grand canonical ensemble, with temperature and
electrostatic potential fixed at the boundary, there is a black hole
solution that is globally stable, which means it dominates the grand
partition function. These results are generally different from the
results obtained for the spherical counterpart of this black hole, the
Reissner-Nordstr\"om-anti-de~Sitter black hole, for which there were
found one or two solutions, that can be stable or unstable, and do not
necessarily dominate the grand partition function
\cite{rnads,Louko}. This means that, contrary to the
Reissner-Nordstr\"om-anti de~Sitter black hole, for the toroidal black
hole  no phase transition was found.

\section*{Acknowledgments}

While this work was in progress the paper by D. R. Brill, J. Louko and
P. Peld\'an  has appeared in the archives, gr-qc/9705012, and has now
been published, see \cite{Brill}. Some of our results intercept and
agree with theirs.  

C.S.P. acknowledges a research grant from JNICT FMRH/BIC/1535/95.


\end{document}